\documentclass{elsart}
\usepackage{epsfig}
\usepackage{rotate}
\usepackage{amssymb}
\usepackage{amsmath}

\newcommand\cc{{\rm \ CC}}

\newcommand\numu{{\nu_\mu}}

\newcommand{\vo}{\mbox{$V^0$}}
\newcommand{\lam}{\mbox{$\rm \Lambda$}}
\newcommand{\lamdecay}{\mbox{$\rm \Lambda \to p \pi^-$}}
\newcommand{\alam}{\mbox{$\rm \bar \Lambda$}}
\newcommand{\alamdecay}{\mbox{$\rm \bar \Lambda \to \bar p \pi^+$}}
\newcommand{\ko}{\mbox{$\rm K^0_s$}}
\newcommand{\kodecay}{\mbox{$\rm K^0_s \to \pi^+ \pi^-$}}

\newcommand{\gamconver}{\mbox{$\rm \gamma \to e^+ e^-$}}

\begin{document}
\begin{frontmatter}
%\centerline{EUROPEAN LABORATORY FOR PARTICLE PHYSICS}
%\bigskip
%\rightline{CERN-EP/99-??}
%\rightline{?? , 1999}
%\rightline(DRAFT, September 17, 1999}
\title{\boldmath 
Measurement of the $\alam$ Polarization \\
in $\nu_\mu$ Charged Current Interactions \\
in the NOMAD Experiment}
%\collab{NOMAD Collaboration}
%\bigskip
\centerline{\bf NOMAD Collaboration}
\vskip 0.5cm

\author[Paris]             {P.~Astier}
\author[CERN]              {D.~Autiero}
\author[Saclay]            {A.~Baldisseri}
\author[Padova]            {M.~Baldo-Ceolin}
\author[Paris]             {M.~Banner}
\author[LAPP]              {G.~Bassompierre}
\author[Saclay]            {N.~Besson}
\author[CERN,Lausanne]     {I.~Bird}
\author[Johns Hopkins]     {B.~Blumenfeld}
\author[Padova]            {F.~Bobisut}
% ...

\address[LAPP]           {LAPP, Annecy, France}                               
\address[Johns Hopkins]  {Johns Hopkins Univ., Baltimore, MD, USA}            
% ...
\address[CERN]           {CERN, Geneva, Switzerland}                          
\address[Lausanne]       {University of Lausanne, Lausanne, Switzerland}      
% ...
\address[Padova]         {Univ. of Padova and INFN, Padova, Italy}            
\address[Paris]          {LPNHE, Univ. of Paris VI and VII, Paris, France}    
% ...
\address[Saclay]         {DAPNIA, CEA Saclay, France}                         
%\address[ANSTO]          {ANSTO Sydney, Menai, Australia}
% ...

\thanks[Perugia]         {Now at Univ. of Perugia and INFN, Perugia, Italy}

\clearpage
\begin{abstract}
We present a measurement of the polarization of 
$\alam$ hyperons produced in $\nu_\mu$ charged current interactions. 
The full data sample from the NOMAD experiment 
has been analyzed using the same $\vo$ identification procedure 
and analysis method reported in a previous paper~\cite{NOMAD-polar}
for the case of $\lam$ hyperons. 
The $\alam$ polarization has been measured for the first time in a neutrino
experiment. 
The polarization vector is found to be compatible with zero. 

\end{abstract}
%\PACS 13.15.+g
\begin{keyword} 
neutrino interactions, antilambda polarization, nucleon spin, 
spin transfer
\end{keyword}
\end{frontmatter}

%%% introduction %%%
\section{Introduction}

The spin structure of hadrons has been extensively studied both experimentally
and theoretically over the past two decades. Measuring parton spin distributions
in different octet baryons can shed light on many phenomena in non-perturbative QCD,
such as $SU(3)$ symmetry breaking, baryon spin content, 
flavor asymmetry in the baryon sea,
which were evoked 
in connection with the proton spin puzzle~\cite{ProtonPuzzle}.
Despite some achievements in this field, our current knowledge of the nucleon 
sea-quark spin distributions and the spin content of the other hyperons 
is still very poor. 
One way to learn about the quark spin distributions of unstable hadrons is to
measure 
the polarized
quark fragmentation function $\Delta D^h_q(z)$ 
of the quark $q$ into the hadron $h$, 
where $z$ is the fraction of the total hadronic energy carried by 
the hadron in the laboratory system. 

Among various hadrons produced in deep inelastic scattering (DIS) 
$\lam$ and $\alam$ hyperons are unique 
because of
their 
parity violating decays into a pair of charged hadrons ($\lamdecay$ and 
$\alamdecay$ respectively) which can be efficiently reconstructed 
and identified.

The $\lam$ ($\alam$)
polarization is measured by the asymmetry 
in the angular distribution of the protons (antiprotons) or pions in the 
parity violating decay process $\lamdecay$ ($\alamdecay$).
In the $\lam$ ($\alam$) rest frame the angular distribution of the
decay protons 
(pions) 
is given by:
\begin{equation}
\frac{1}{N}\frac{\d N}{\d \Omega} = \frac{1}{4\pi}(1+\alpha \mathbf P 
\cdot \mathbf k),
\label{eq:asymmetry}
\end{equation}
where $\mathbf P$ is the $\lam$ ($\alam$) polarization vector,
$\alpha = 0.642 \pm 0.013$~\cite{PDG} 
is the decay asymmetry parameter 
and $\mathbf k$ is the unit vector along the direction of the outgoing positive
decay particle (the proton in case of $\lam$ and the $\pi^+$ in case of $\alam$).

To probe the polarized quark distribution in a hadron $\Delta q_h(x_{Bj})$,
where $x_{Bj}$ is the standard Bjorken variable,
a source of polarized quarks is required. 
Of the 
many means of obtaining
polarized quarks~\cite{Zpole,charged},
neutrino and antineutrino DIS are exceptional due to 
their 100\% natural polarization. Moreover, weak interactions provide a 
{\it source of polarized quarks of specific flavour}, which makes 
the measurement of $\lam$ and $\alam$ polarizations
in (anti)neutrino DIS an ideal tool to investigate 
different spin transfer mechanisms and to check various models of 
the baryon spin content. 

Different physical mechanisms are responsible for the $\lam$ and $\alam$ 
polarization in the fragmentation regions defined by positive and negative
values of $x_F = 2 p^*_L/W$.
A polarization of the strange sea in the nucleon~\cite{Ellis} can manifest itself 
through
a polarization of $\lam$ and $\alam$ hyperons produced in the 
target fragmentation region ($x_F < 0$) in (anti)neutrino DIS 
process.
Measurements of the polarization
of the $\lam$ and $\alam$ hyperons produced in the 
current fragmentation region ($x_F > 0$) in (anti)neutrino DIS can provide 
information on the 
polarized
fragmentation functions
of quarks and antiquarks into a given hyperon 
($h = \lam$ or $\alam$)~\cite{Ma-Soffer,Ma2000,Ma_et_al}. 
The longitudinal polarization of a hyperon $h$ produced 
in neutrino DIS in the current fragmentation 
region, assuming no polarization transfer from the other
(anti)quarks fragmenting into this hyperon, is given by:
{\small
\begin{equation}
P^{h}_{\nu} = - 
\frac{\left[ d(x_{Bj}) + \omega s(x_{Bj})\right] \Delta D_u^{h}(z) - (1-y_{Bj})^2 \bar u(x_{Bj})  
\left[ \Delta D_{\bar d}^{h}(z) + \omega  \Delta D_{\bar s}^{h}(z) \right]}
{\left[d(x_{Bj}) + \omega s(x_{Bj})\right] D_u^{h}(z) + (1-y_{Bj})^2 \bar u(x_{Bj})  \left[ D_{\bar d}^{h}(z) + 
\omega  D_{\bar s}^{h}(z) \right]},
\label{lampolar}
\end{equation}
}
where $\omega = \tan^2\theta_C$ ($\theta_C$ is the Cabibbo angle).
A measurement of the $\lam$ polarization in the current fragmentation region 
in $\numu$ charged current (CC) DIS 
provides an estimate of the spin transfer coefficient
$C_u^{\lam} = \Delta D_u^{\lam}(z)/D_u^{\lam}(z)$ 
because of the dominant contribution from the first term in both the numerator and denominator of Eq.~(\ref{lampolar}).
However an interpretation of the $\alam$ polarization in
the current fragmentation region in $\numu \cc$ DIS is more complicated 
since the two terms in both the numerator and denominator of Eq.~(\ref{lampolar}) could be of comparable size.

In order
to reduce the number of independent fragmentation functions 
the authors of~\cite{Ma-Soffer} made the following assumptions: 
\begin{equation}
\label{symmetry-nonpol}
D_q^{\lam}(z) = D_u^{\lam}(z) = D_d^{\lam}(z) = D_{\bar u}^{\alam}(z) =
D_{\bar d}^{\alam}(z)
\end{equation}
and 
\begin{equation}
\label{symmetry-pol}
\Delta D_q^{\lam}(z) = \Delta D_u^{\lam}(z) = \Delta D_d^{\lam}(z) = 
\Delta D_{\bar u}^{\alam}(z) = \Delta D_{\bar d}^{\alam}(z).
\end{equation}
The relations~(\ref{symmetry-nonpol}) and~(\ref{symmetry-pol})
simplify
the interpretation of the results of the $\lam$ and
$\alam$ polarization measurements in (anti)neutrino DIS.
However, these relations may not be valid 
if
a sizable fraction of 
$\lam$ and $\alam$ hyperons is produced via resonance 
or heavier hyperon decays.

The NOMAD experiment~\cite{NOMAD} 
has collected $1.3 \times 10^6$ $\nu_\mu$ CC events 
and has observed 
an unprecedented number
of $\lam$ and $\alam$ decays. 
A detailed paper devoted to the $\lam$ polarization
measurement in the NOMAD experiment has been  
published recently~\cite{NOMAD-polar}. 
We 
present 
here
the
first measurement 
of the polarization of 
$\alam$ hyperons produced in  $\nu_\mu$ CC interactions.

%%% Event Selection %%%
\section{Event selection and $\vo$ identification}

The 
measurement
of charged tracks produced in neutrino interactions
in the NOMAD detector 
is
performed with 
a set of drift chambers~\cite{nomad_dc} 
located inside a dipole magnet 
with
a field of 0.4 Tesla. 

\begin{figure}[htb]
\begin{center}
\vspace*{-6cm}
\epsfig{file=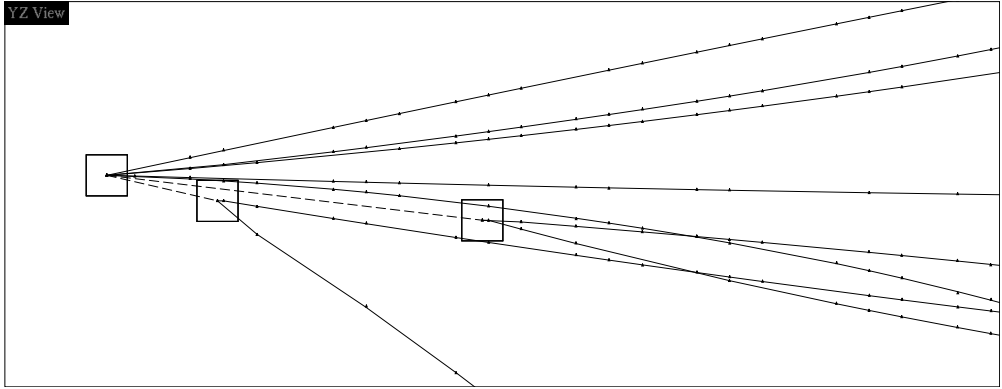,angle=-90,width=150mm}
\vspace*{-7cm}                                                                  
\protect\caption{\it A reconstructed data event containing 
2 $\vo$ vertices identified as $\lam$ and $\alam$
decays by 
our
identification procedure.
The scale on this plot is given by the size of the vertex boxes  
($3\times3$ cm$^2$).
}
\label{fig:lambda-antilambda}
\end{center}
\end{figure}

A $\alam$ decay appears in the detector as a $\vo$-like vertex: 
two tracks of opposite charge
emerging from a common vertex separated from the primary neutrino interaction 
vertex (see Fig.~\ref{fig:lambda-antilambda}).
A $\vo$-like signature is expected also for $\lam$ and $\ko$ decays and for photon conversions. 

The data are compared to the results
of a Monte Carlo (MC) simulation based on 
LEPTO 6.1~\cite{LEPTO} and
JETSET 7.4~\cite{JETSET} generators for neutrino
interactions and on a GEANT~\cite{GEANT} based program for the detector 
response. 
In our MC simulation $\alam$ hyperons are not polarized.

The selection procedure for the $\numu \cc$ event sample used in this analysis
has been described in~\cite{NOMAD-polar}. 

Since the NOMAD detector is unable to distinguish (anti)protons 
from pions in the momentum range relevant to this analysis,
our $\vo$ identification procedure relies on the kinematic properties of 
a $\vo$ decay.

For the $\vo$ identification a kinematic fit method has been used 
as described in reference~\cite{NOMAD-polar}.
This fit has been performed for three decay hypotheses: 
$\kodecay$, 
$\lamdecay$, $\alamdecay$ and for 
the hypothesis of a photon conversion 
$\gamconver$.
The output of the kinematic fits applied to a given $\vo$ vertex
consists of four $\chi^2_{V^0}$.
The different
regions in the four-dimensional $\chi^2_{V^0}$ space
populated by particles identified as $\lam$, $\alam$
and $\ko$
have been selected. 

Identified $\vo$'s 
are
of two types:
\begin{itemize}
\item {\em uniquely} identified $\vo$'s, which, in 
the four-dimensional $\chi^2_{V^0}$ space described above, populate regions 
corresponding 
to the decay of different particles;
\item {\em ambiguously} identified $\vo$'s, which populate overlapping 
kinematic regions 
where
the decays of different particles
are simultaneously present.
\end{itemize}
The treatment of ambiguities aims at selecting
a given $\vo$ decay with the highest efficiency 
and the lowest background contamination from other $\vo$ types.
Our identification strategy consists of two steps:
\begin{itemize}
\item[1)] we select a sample of uniquely identified 
$\alam$ hyperons which has a purity 
of
91\%;
\item[2)] we allow ourselves an additional 7\% contribution from a 
subsample of ambiguously identified $\alam$ particles resolving
the ambiguities between 
$\alam$ and $\ko$ 
in favour of
maximal purity.
\end{itemize}
This approach provides an optimum compromise between high statistics 
and well understood background contamination.

\begin{figure}[htbp]
\begin{center}
\begin{tabular}{cc}
\epsfig{file=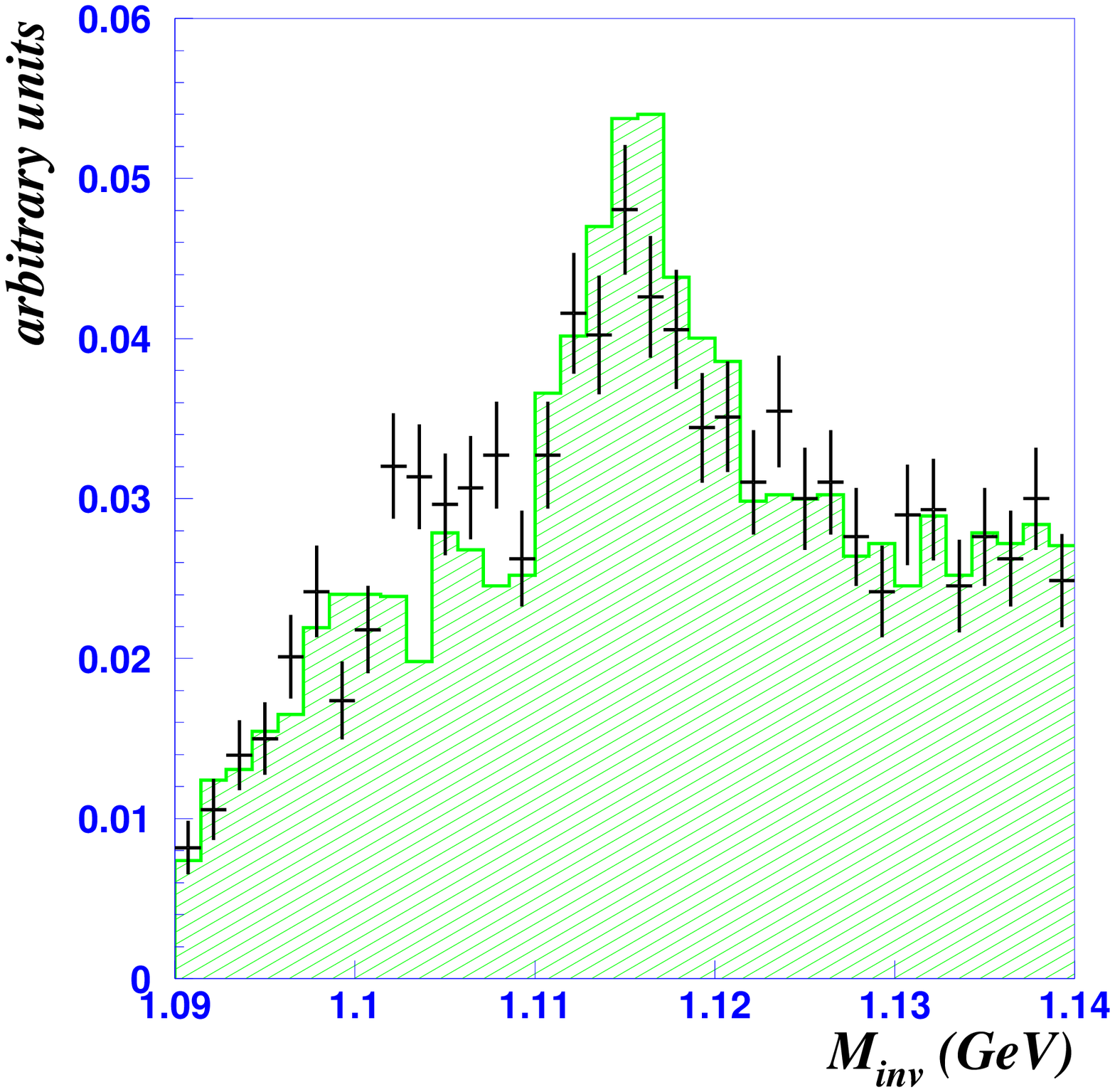,width=0.45\linewidth}&
\epsfig{file=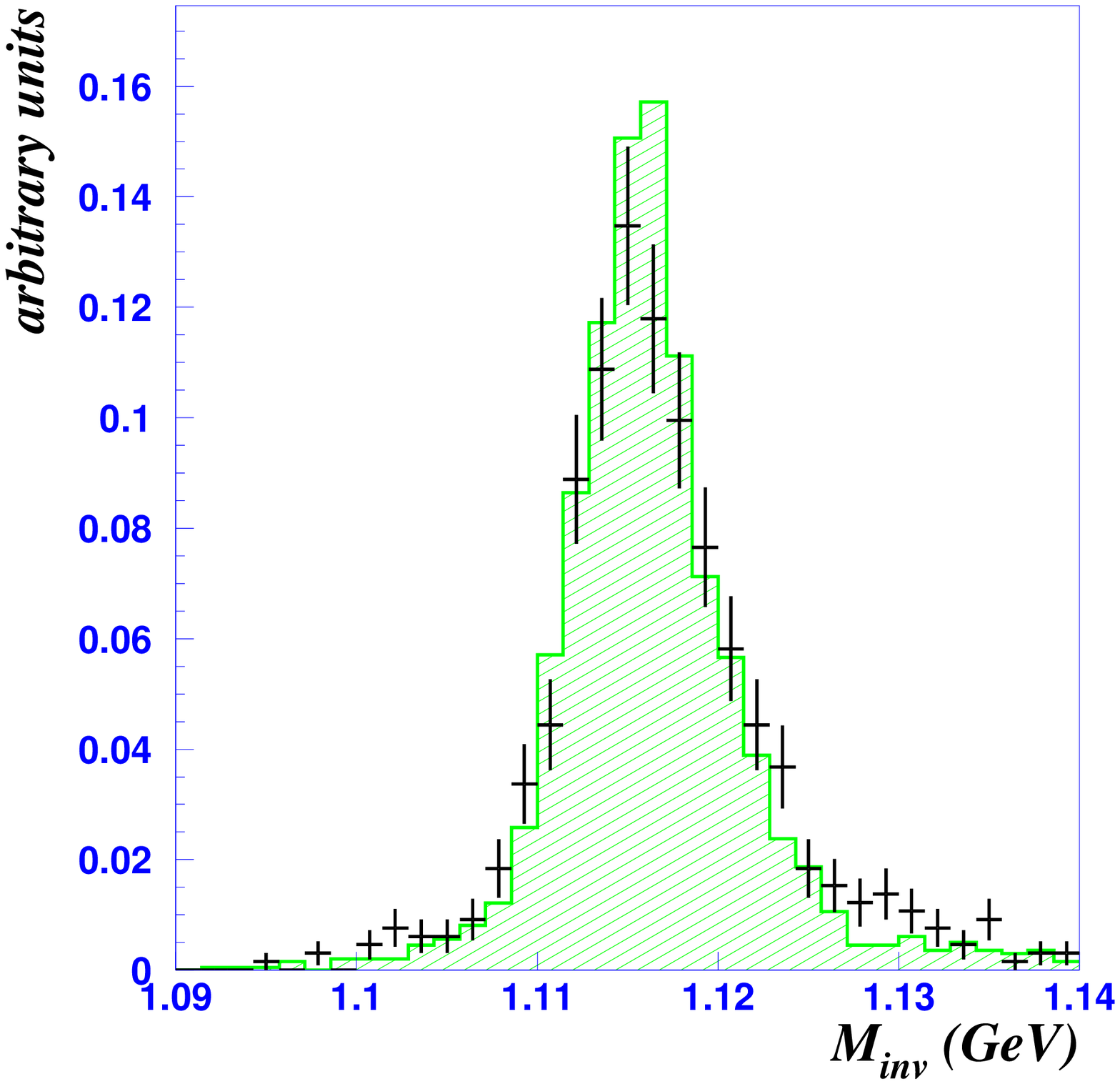,width=0.45\linewidth}\\
\end{tabular}
\end{center}
\protect\caption{\it Normalized invariant mass distributions in both data (points with error bars) 
and MC (histogram) calculated for $\vo$ vertices under the assumption of a $\alam$ decay before (left) and after 
(right) the $\vo$ identification procedure.}
\label{fig:ident-mass-antilambda}
\end{figure}

A 
MC simulation program
has been used to study the purity of the overall $\alam$ sample obtained
by our selection criteria 
in $\nu_\mu$ CC interactions.
This study indicates that
89.6\% of the selected $\vo$'s are true $\alam$ hyperons, 
5.4\% are misidentified $\ko$ mesons and 5.0\% 
are due to
random track associations. 
The global reconstruction and identification efficiency 
is 18.6\%.
A total of 649 identified $\alam$ decays is found in our data, 
representing  
a significantly
larger 
number
than in
previous (anti)neutrino experiments performed 
with bubble chambers~\cite{ammosov,allasia,Jones,Willocq,Prospo}.
Fig.~\ref{fig:ident-mass-antilambda} shows the invariant mass distributions for
$\bar p \pi^+$ combinations
before and after the $\vo$ identification procedure. 

\begin{figure}[htb]
\begin{center} 
\epsfig{file=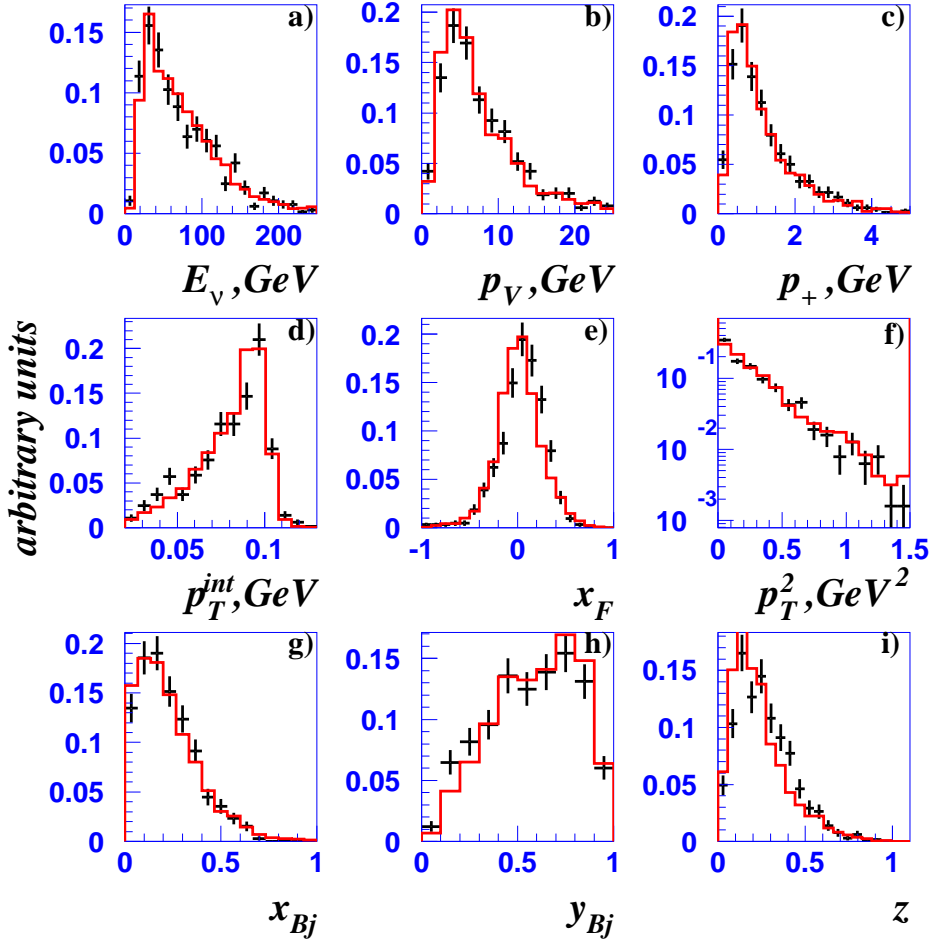, width=1.0\linewidth
}
\end{center}
\caption{\it Comparison of reconstructed kinematic variables for $\nu_\mu$ CC
events containing identified $\alam$ hyperons for simulated (histogram) and
data (points with error bars) events: 
a) neutrino energy, b) $\alam$ momentum,
c) momentum of the outgoing $\pi^+$ from the $\alam$ decay,
d) transverse component of the momentum of one of the 
outgoing charged tracks with respect to the $\vo$ 
direction
(internal $p_T$), 
e) $x$-Feynman, 
f) square of the transverse component of the $\alam$ momentum 
with respect to the hadronic jet direction, g) $x$-Bjorken, h) $y$-Bjorken,
i) fraction of the total hadronic energy carried by the $\alam$ 
in the laboratory system.
}
\label{fig:variables}
\end{figure}

We have compared the distributions of many kinematical variables, describing
the neutrino interaction ($E_\nu$, $x_{Bj}$, $y_{Bj}$), $\vo$ production and
decay ($p_V$, $p_+$, $p^{int}_T$), $\vo$ behaviour in the hadronic jet
($p_T$, $x_F$, $z$), obtained in data and MC events. 
The comparison (see
Fig.~\ref{fig:variables}) shows that the experimental distributions 
of all these kinematical variables 
are in 
agreement with the MC results.
However, we note that the $x_F$ distribution is shifted towards positive 
values in the data (see Fig.~\ref{fig:variables}e), while it is centered
around
zero in the MC. 
A
small discrepancy between data and MC
may be noticed in the $z$ distribution (Fig.~\ref{fig:variables}i).

%%% Analysis And Systematics %%%
\section{Polarization analysis}
For the polarization analysis described below 
we use the ``J'' reference system, in which the axes are defined as follows 
(in the $\alam$ rest frame):
\begin{itemize}
\item the $\mathbf n_x$ axis is chosen along the reconstructed 
$W$-boson direction ($\vec e_W$);
\item the $\mathbf n_y$ axis is orthogonal to the $\alam$ production plane 
(defined as the plane containing both the target nucleon ($\vec e_{T}$)
and the $W$-boson vectors): \\
$\mathbf n_y= \vec e_W \times \vec e_{T} /
|\vec e_W \times \vec e_{T} |$.
\item the $\mathbf n_z$ axis is chosen to form a right-handed coordinate system: \\
$\mathbf n_z = \mathbf n_x \times \mathbf n_y$. 
\end{itemize}

\begin{figure}[htb]
\begin{center} 
\begin{tabular}{ccc}
\epsfig{file=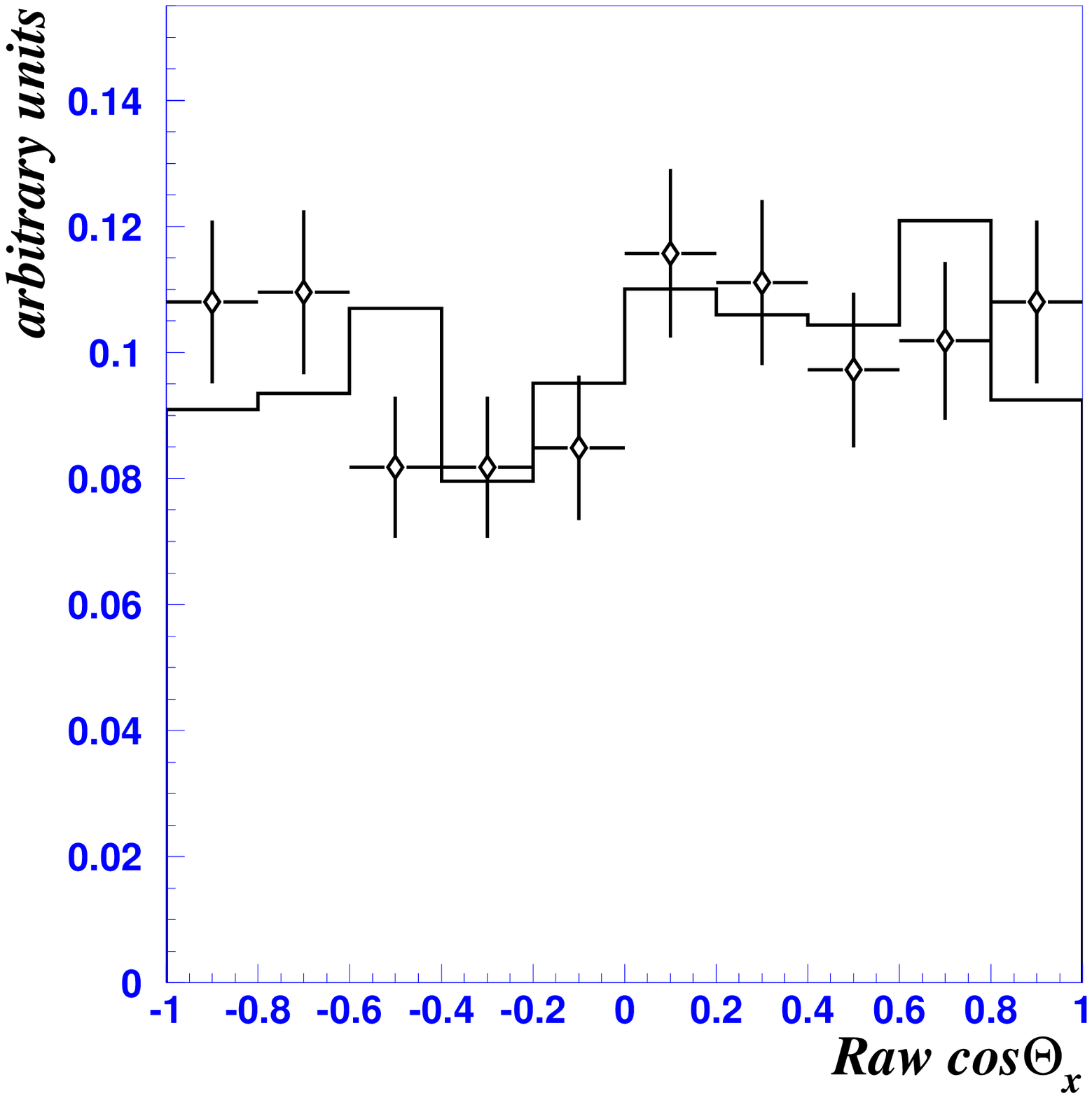, width=0.35\linewidth} &
\epsfig{file=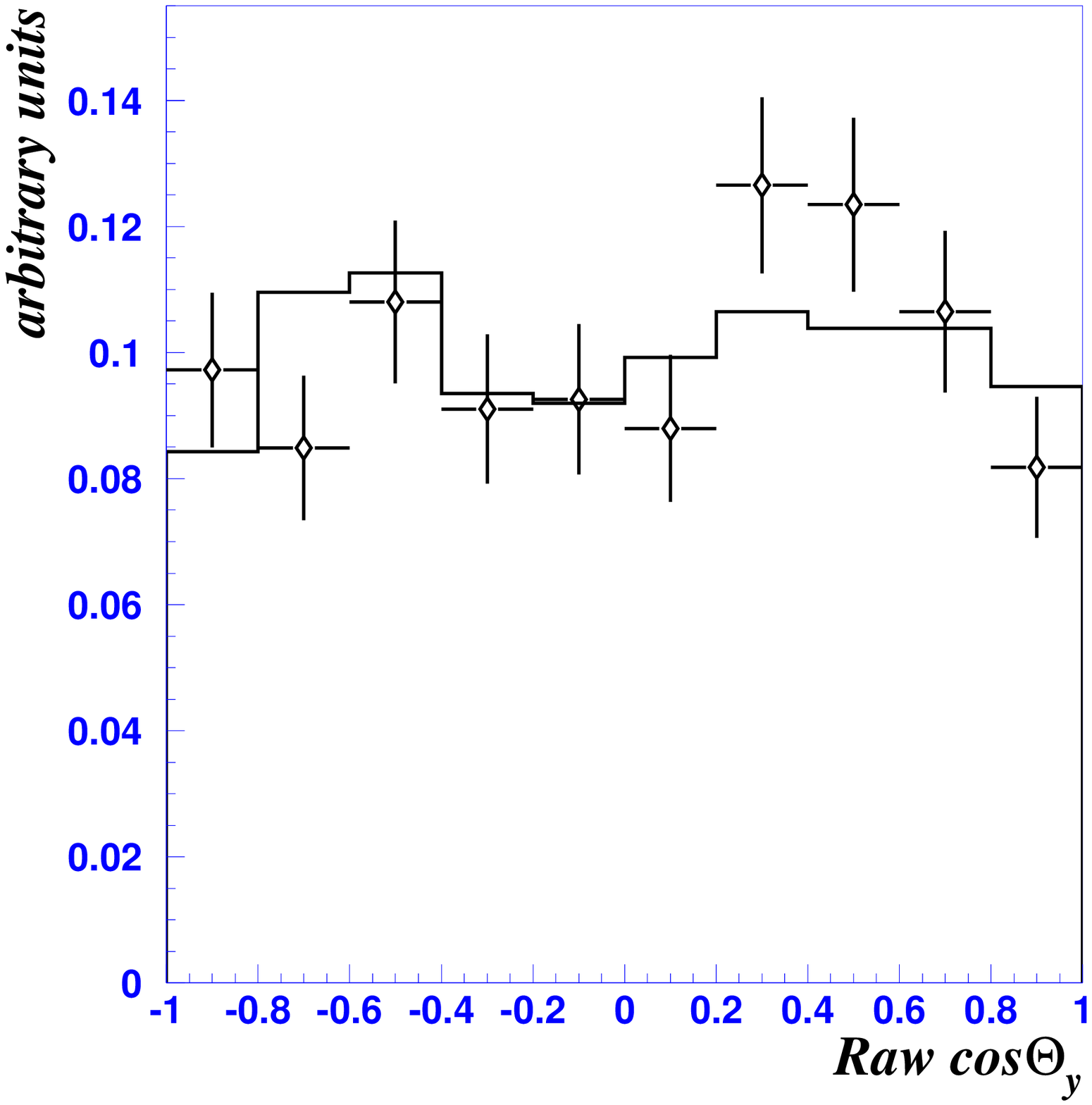, width=0.35\linewidth} &
\epsfig{file=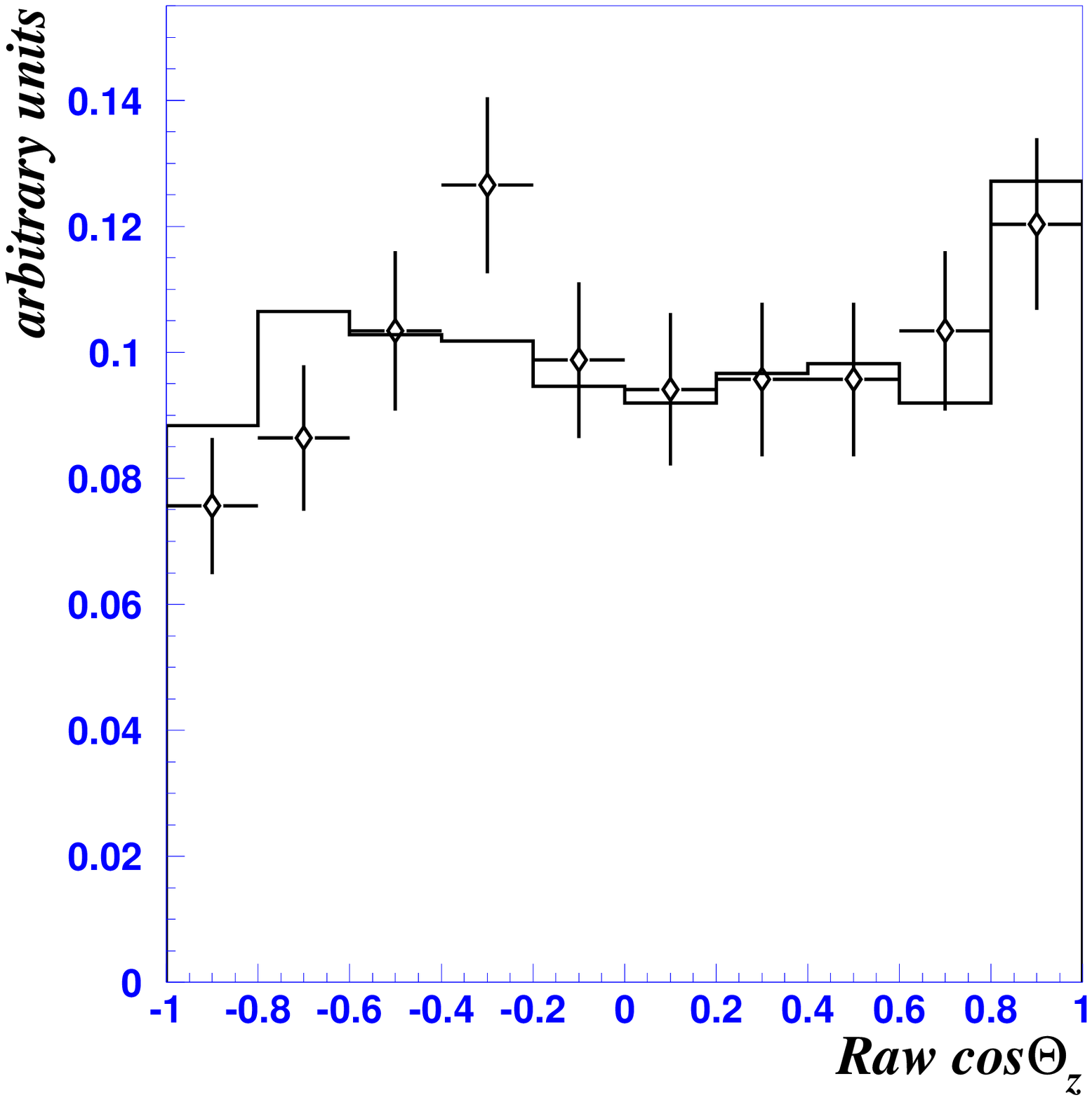, width=0.35\linewidth} \\
\epsfig{file=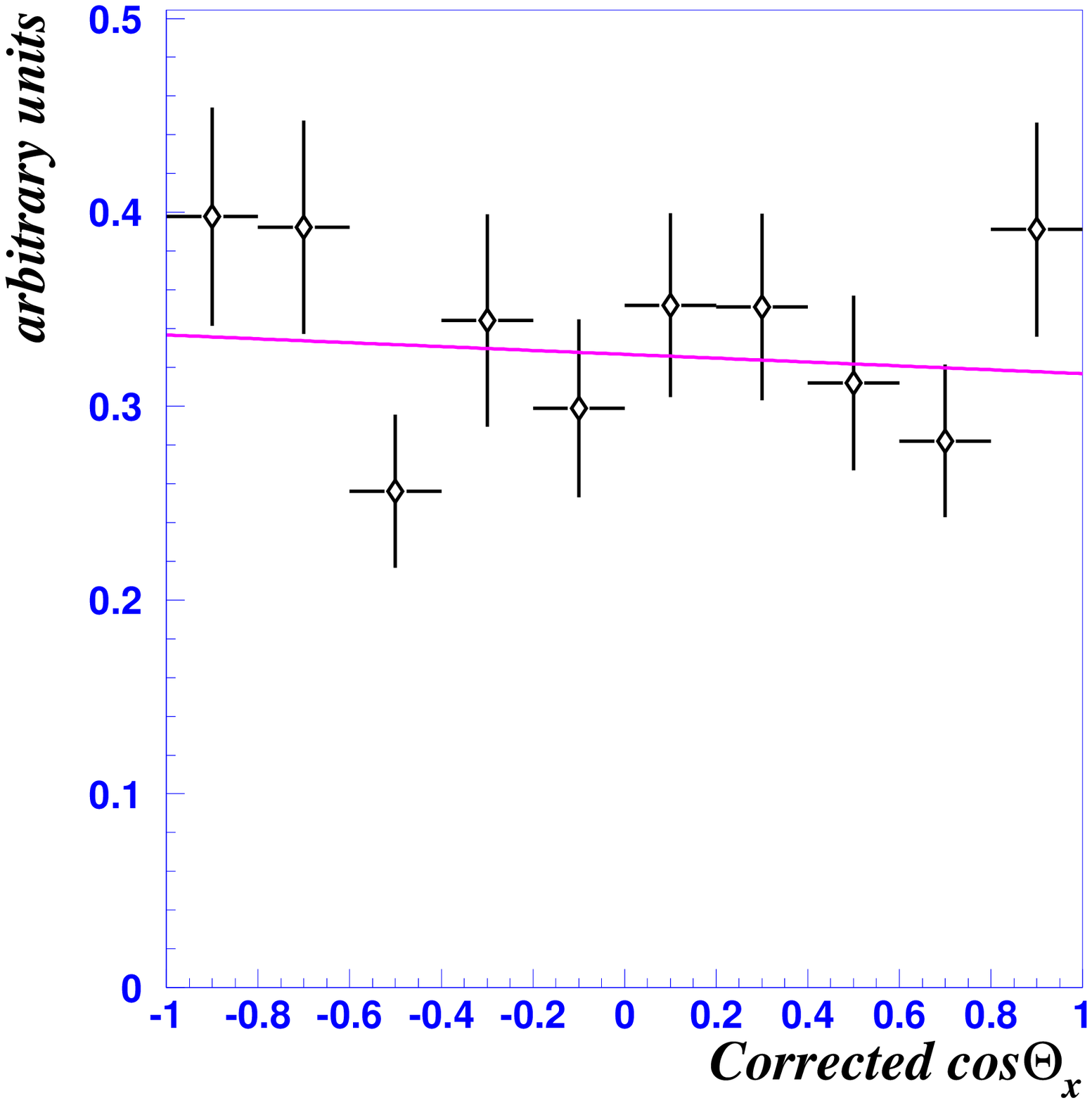, width=0.35\linewidth} &
\epsfig{file=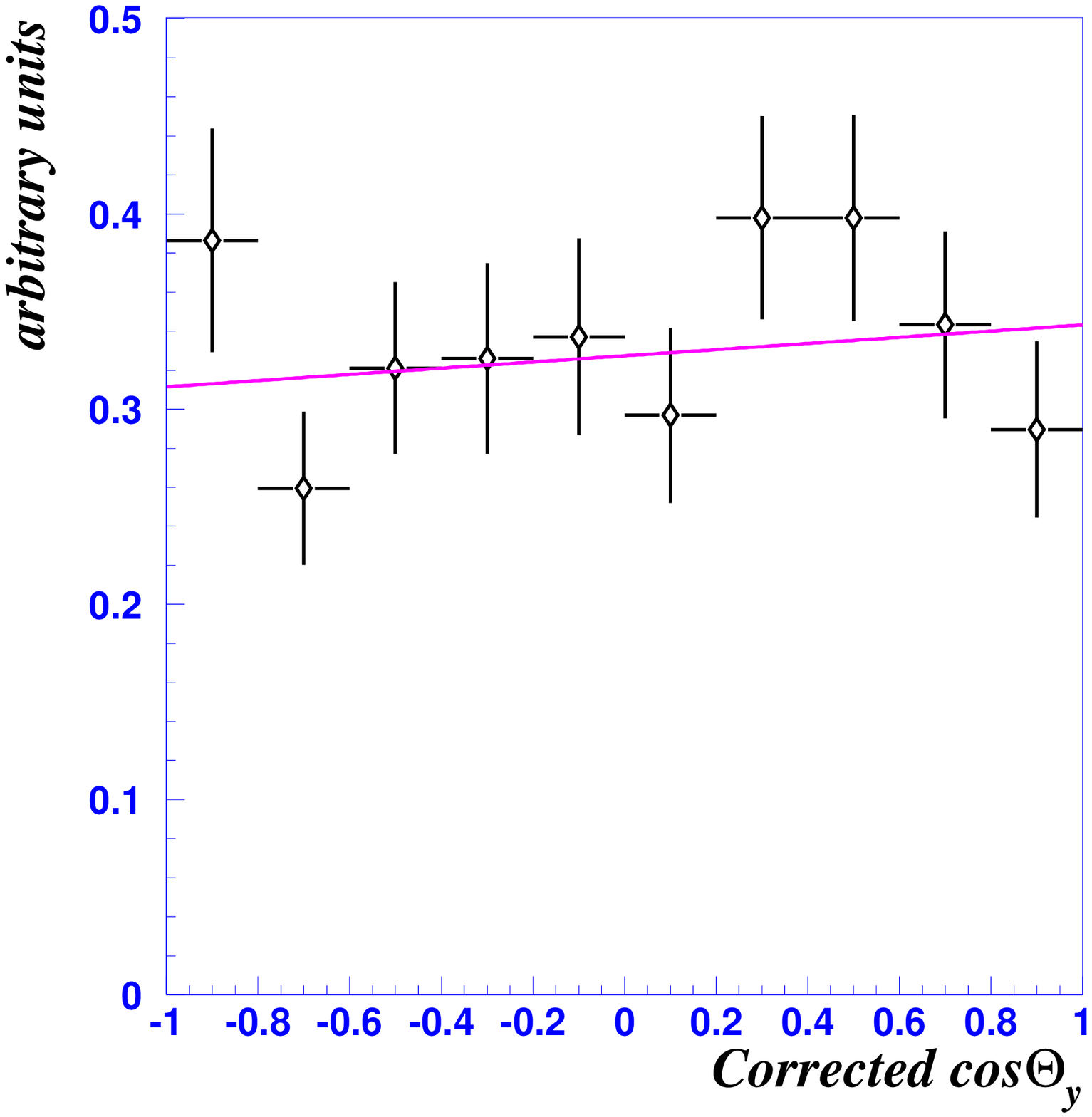, width=0.35\linewidth} &
\epsfig{file=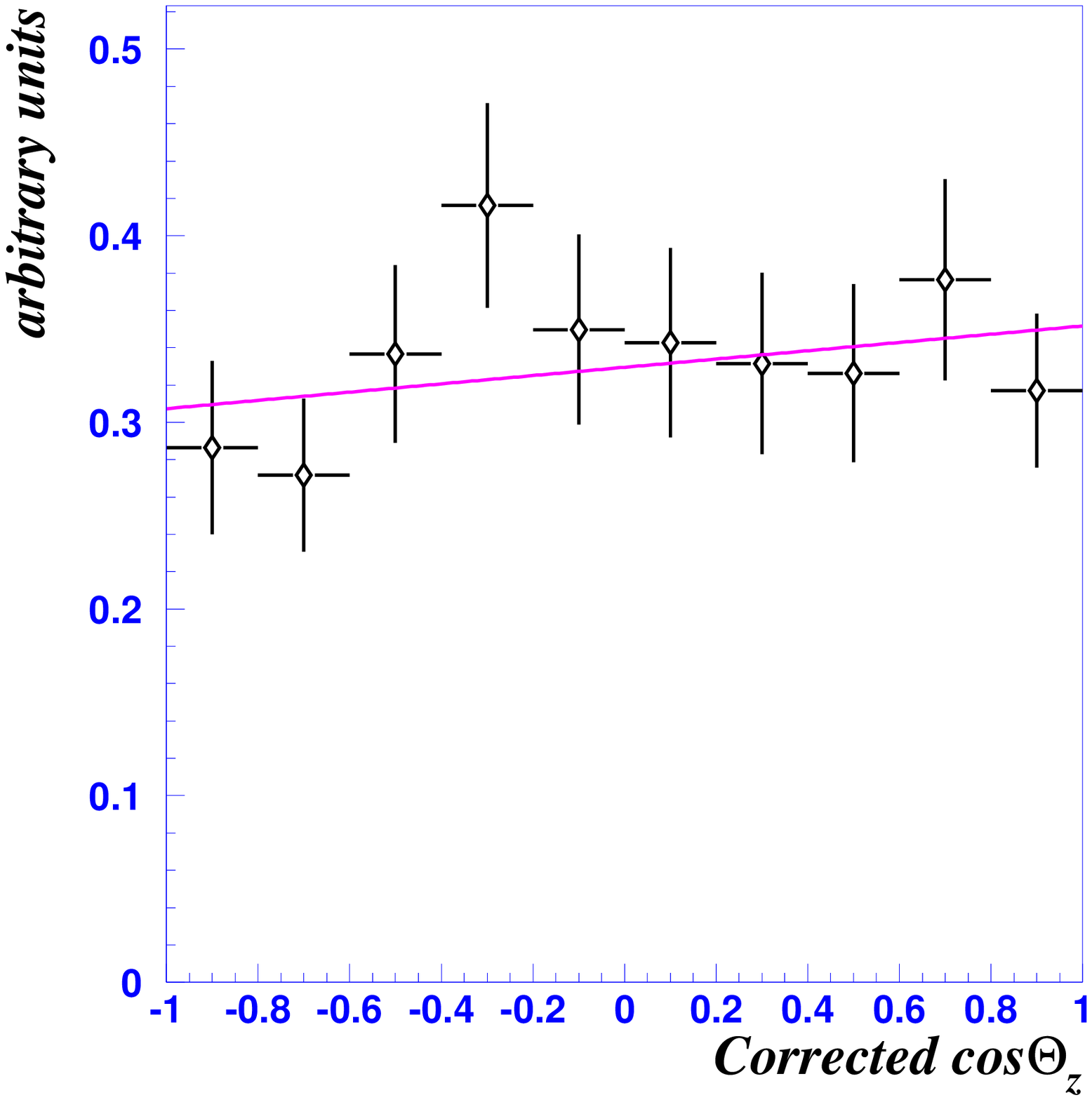, width=0.35\linewidth} \\
\end{tabular}
\end{center}
\caption{\it Top: normalized angular distributions ($\cos \theta_i$)
for $\alam$ hyperons in reconstructed Monte Carlo events 
(histogram) and in data (points with error bars). Bottom: 
experimental angular distributions corrected for detector acceptance and selection effects;
the polarization is given by the slope of the corresponding linear
fit.}
\label{fig:cosines}
\end{figure}

The experimental resolution 
for
the
reconstructed 
$\cos\theta_i = {\mathbf n}_i \cdot \mathbf k$, where $\mathbf k$ is the 
unit vector in the direction of the outgoing positive track ($\pi^+$), 
is found to be about $0.07$. 
The raw $\cos\theta_i$ distributions of $\alam$ hyperons 
are affected by the detector acceptance and the reconstruction algorithm:
\begin{equation}
\label{eq:distorted-distr}
\frac{dN(\cos\theta_i)}{d\cos\theta_i} = A(\cos\theta_i) (1+\alpha P_i \cos\theta_i),
\end{equation}
where $A(\cos\theta_i)$ is the 
detector acceptance
function which also depends on other kinematic 
variables.
The 
angular distributions in the $\alam$ case (see Fig.~\ref{fig:cosines}) are less distorted than 
in the case of 
the $\lam$ sample (see e.g.~\cite{NOMAD-polar}) because of 
a higher average momentum of $\alam$ hyperons produced in
$\numu \cc$ interactions. This is due to baryon number conservation which
implies a higher $W^2$ threshold for $\alam$ production in neutrino-nucleon DIS. 

A method which allows the extraction of all three components of the $\lam$ ($\alam$)
polarization vector at the same time, taking
into account 
both the detector acceptance and the 
differences between generated and reconstructed angular variables, 
has been developed~\cite{NOMAD-polar,memo2000-1}.
The one-dimensional option of this method is used for the analysis presented
here because of the low statistics of the $\alam$ sample. 

The results are summarized in Table~\ref{tab:antilambda-numu}. 

\begin{table}[htb]
\begin{center} 
\caption{\it $\bar \Lambda$ polarization in $\nu_\mu$ CC events 
(the first error is statistical and the second is systematic).}
\begin{tabular}{||c|c|c|c|c||}
\hline
\hline
 & & \multicolumn{3}{|c|}{$\bar \Lambda$ Polarization} \\
\cline{3-5}
Selection  & Entries  &  $P_x$         &  $P_y$         &$P_z$ \\
\hline
\hline
full sample & 649     &$-0.07 \pm 0.12 \pm 0.09$&$0.09 \pm 0.13 \pm 0.10$&$ 0.10 \pm 0.13 \pm 0.07$\\
\hline
\hline
$x_F<0$     & 248     &$ 0.23 \pm 0.20 \pm 0.15$&$0.04 \pm 0.20 \pm 0.19$&$-0.08 \pm 0.21 \pm 0.12$\\
$x_F>0$     & 401     &$-0.23 \pm 0.15 \pm 0.08$&$0.10 \pm 0.17 \pm 0.05$&$ 0.25 \pm 0.16 \pm 0.06$\\
\hline
\hline
$x_{Bj}<0.2$& 331    &$-0.12 \pm 0.17 \pm 0.08$&$ 0.08 \pm 0.18 \pm 0.11$&$ 0.01 \pm 0.17 \pm 0.07$\\
$x_{Bj}>0.2$& 318    &$-0.03 \pm 0.17 \pm 0.14$&$ 0.10 \pm 0.18 \pm 0.10$&$ 0.20 \pm 0.19 \pm 0.07$\\
\hline
\hline
\end{tabular}
\label{tab:antilambda-numu}
\end{center} 
\end{table}

\section{Systematic errors}

The following potential sources of systematic errors have been studied
(see details in~\cite{NOMAD-polar}):
\begin{itemize}
\item uncertainty in the incoming neutrino energy determination
resulting in an uncertainty in the reconstructed W-boson direction and, thus,
leading to a poor definition of the ``J'' reference system 
in which the $\alam$ polarization is measured;
\item 
uncertainty in the
background rate caused by 
possible 
differences between 
MC and data;
\item 
dependence of the final results on the selection criteria; 
\item spin precession 
due to the $\alam$
travelling through the magnetic field of the detector.
\end{itemize}

We summarize the results of this study for the full $\alam$ sample
in Table~\ref{tab:syst_err}.
The maximal deviation with respect to the reference result is used 
as an estimate of the systematic uncertainty.
The overall
systematic errors 
are
obtained
by adding all 
the contributions in quadrature\footnote{Neglecting possible correlations between variations
of selection criteria and resulting background uncertainties.}.
In the $\alam$ analysis the statistical errors are larger than
the systematic errors.

\begin{table}
\begin{center}
\caption{\it Summary of relative systematic errors 
on the three components of the $\alam$ polarization vector for the full $\alam$ sample.
}
\label{tab:syst_err}
\begin{tabular}{||c|c|c|c|c|c||}
\hline
\hline
$P_i$& $\nu$ energy   & background & variation of & $\alam$ spin  & total  \\
     & reconstruction & uncertainties    & selection criteria   & precession & \\
\hline
$P_x$&$2.9\cdot10^{-2}$&$5.1\cdot10^{-2}$&$6.9\cdot10^{-2}$&$9.1\cdot10^{-3}$&$9.1\cdot10^{-2}$\\
$P_y$&$1.0\cdot10^{-1}$&$9.9\cdot10^{-3}$&$2.1\cdot10^{-2}$&$4.9\cdot10^{-3}$&$1.0\cdot10^{-1}$\\
$P_z$&$7.4\cdot10^{-3}$&$3.7\cdot10^{-2}$&$5.3\cdot10^{-2}$&$3.3\cdot10^{-3}$& $6.5\cdot10^{-2}$\\
\hline
\hline
\end{tabular}
\end{center}
\end{table}

%%% Results %%%
\section{
%Results and 
%d
Discussion}

In our previous publication~\cite{NOMAD-polar} we have reported the
measurements of the $\lam$ polarization  in $\nu_\mu \cc$ DIS
obtained 
from
a sample of 8087 
identified 
$\lamdecay$ decays. Let us first briefly recall these results.

For the longitudinal polarization we have found:
\begin{itemize}
\item in the target fragmentation region
$P_x^{\lam} = -0.21 \pm 0.04 \mbox{(stat)} \pm 0.02 \mbox{(syst)}$.\\
This result is in qualitative agreement with the predictions of the
model of negatively polarized $s \bar s$ pairs in the nucleon~\cite{Ellis}.
\item in the current fragmentation region
$P_x^{\lam} = -0.09 \pm 0.06 \mbox{(stat)} \pm 0.03 \mbox{(syst)}$.\\
This result provides a measure of the spin transfer coefficient 
$C_u^{\lam} \approx - P_x^{\lam}$ (see Eq.~(\ref{lampolar})).
\end{itemize}
A significant transverse polarization 
$P_y^{\lam} = -0.22 \pm 0.03 \mbox{(stat)} \pm 0.01 \mbox{(syst)}$
has also been observed. 
Its dependence on the transverse momentum of the $\lam$ with 
respect to the hadronic jet direction is in qualitative
agreement with the well established behaviour observed in unpolarized 
hadron-hadron experiments~\cite{Felix}.

As shown in Table~\ref{tab:antilambda-numu}, 
the observed $\alam$ polarization vector for the full data sample
is consistent with zero. 
If we split the $\alam$ 
sample
into subsamples with $x_F > 0$ and $x_F < 0$, 
we still find results consistent with zero.
However, 
a 
negative
value of $P_x^{\alam} (x_F>0)$, if confirmed,
would disagree with the expectations 
based on calculations~\cite{Kotzinian} 
performed in the framework of the naive quark model, as well as of
the Burkardt-Jaffe~\cite{BJ} and 
the Bigi-Gustafson-H\"akkinen models~\cite{Bigi,GH}.
Similarly, a positive value of the $\alam$ longitudinal polarization 
at $x_F<0$,
if confirmed,
would not be
consistent with naive expectations in the framework of the model of
negatively polarized $s\bar s$ pairs in the nucleon~\cite{Ellis}.

No significant transverse polarization has been found for the $\alam$ sample.

As a cross-check we have measured the $\lam$ ($\alam$) transverse polarization using a modification
of  a bias cancelling technique adopted in hadron experiments~\cite{polar-method-hadron}. 
We exploit the left-right symmetry of our 
detector with respect to neutrino beam axis, and separately measure the distributions of Eq.~(\ref{eq:distorted-distr})
for events with primary vertex on the left (L) or on the right (R) of this axis, replacing $\cos\theta_y$ by $-\cos\theta_y$
for the right half of the detector:
\begin{center}
\begin{eqnarray}
&L& = \frac{dN_L}{d\cos\theta_y} = A_L(\cos\theta_y)(1 + \alpha P_y \cos\theta_y)\\
&R& = \frac{dN_R}{d\cos\theta_y} = A_R(-\cos\theta_y)(1 - \alpha P_y \cos\theta_y)
\end{eqnarray}
\end{center}
Using the fact that $A_L(\cos\theta_y)  = A_R(-\cos\theta_y)$, we define an asymmetry
$$
\epsilon = \frac{L-R}{L+R} = \alpha P_y \cos\theta_y
$$
which does not depend on the detector acceptance. By fitting $\epsilon$ to a straight line in $\cos\theta_y$
we extract the transverse polarization $P_y$.

All the measurements of the transverse polarization performed with 
this method are in a good agreement with the results obtained using 
our standard method
for both $\lam$~\cite{NOMAD-polar} and $\alam$~(Table~\ref{tab:antilambda-numu}) samples.

\section{Conclusion}

The results of the $\alam$ polarization measurements 
in $\nu_\mu$ CC DIS in the NOMAD experiment
have been
presented. 
A clean $\alam$ sample has been selected on the 
basis
of kinematic fits
performed 
on
$\vo$-like decays. 
The method used to extract the three components of the $\alam$ polarization 
vector accounts for the smearing of the angular variables.
While the 
statistics of 
the $\alam$ sample is limited to 649 events, 
the $\alam$ polarization has been measured
for the first time in a neutrino
experiment.
The results for the three components of the $\alam$ polarization vector
as measured in the ``J'' reference system are compatible with zero:
$P_x = -0.07 \pm 0.12 \mbox{(stat)} \pm 0.09 \mbox{(syst)}$,
$P_y = 0.09 \pm 0.13 \mbox{(stat)} \pm 0.10 \mbox{(syst)}$,
$P_z = 0.10 \pm 0.13 \mbox{(stat)} \pm 0.07 \mbox{(syst)}$.
In addition
no evidence for the $\alam$ polarization is 
found either in the current or in the target fragmentation regions.

More precise measurements of the $\alam$ polarization in (anti)neutrino 
interactions could help in
clarifying the fragmentation and spin transfer 
mechanisms
of quarks and (anti)quarks into
a bound system of antiquarks.
Improved 
experimental results on the $\alam$ polarization 
in $\nu$($\bar \nu$) DIS will be hard to obtain before 
more intense neutrino beams~\cite{nufact} become
available.

\vspace{.5 cm}
{\large \bf Acknowledgements}

\vspace{.5cm}
We gratefully acknowledge  the CERN SPS accelerator and beam-line staff
for the magnificent performance of the neutrino beam.\
The experiment was  supported by  the following
funding agencies:
Australian Research Council (ARC) and Department of Industry, Science, and
Resources (DISR), Australia;
Institut National de Physique Nucl\'eaire et Physique des Particules (IN2P3), 
Commissariat \`a l'Energie Atomique (CEA),  France;
Bundesministerium f\"ur Bildung und Forschung (BMBF, contract 05 6DO52), 
Germany; 
Istituto Nazionale di Fisica Nucleare (INFN), Italy;
Joint Institute for Nuclear Research and 
Institute for Nuclear Research of the Russian Academy of Sciences, Russia; 
Fonds National Suisse de la Recherche Scientifique, Switzerland;
Department of Energy, National Science Foundation (grant PHY-9526278), 
the Sloan and the Cottrell Foundations, USA.

%%% References %%%

\end{document}